\documentclass[twocolumn,english,aps,pra,10pt,superscriptaddress,floatfix]{revtex4-2}

\usepackage{times}
\usepackage{bm}
\usepackage{graphicx}
\usepackage{amssymb,amsfonts,amsmath,amsbsy,bm,t1enc,latexsym}
\usepackage{float}
\usepackage[colorlinks=true,citecolor=blue,linkcolor=magenta]{hyperref}
\usepackage[english]{babel}
\usepackage{url}%

\usepackage{color,soul}
\usepackage{textcomp}
\usepackage[]{lipsum}

\usepackage[markup=blue]{changes}

\newcommand{\SiN}[0]{Si$_3$N$_4$}

\begin{document}
	
\title{An ultra-broadband photonic-chip-based traveling-wave parametric amplifier} 

\author{Nikolai Kuznetsov}
\thanks{These authors contributed equally to this work.}
\affiliation{Institute of Physics, Swiss Federal Institute of Technology Lausanne (EPFL), CH-1015 Lausanne, Switzerland}
\affiliation{Center of Quantum Science and Engineering (EPFL), CH-1015 Lausanne, Switzerland}

\author{Alberto Nardi}
\thanks{These authors contributed equally to this work.}
\affiliation{Institute of Physics, Swiss Federal Institute of Technology Lausanne (EPFL), CH-1015 Lausanne, Switzerland}
\affiliation{IBM Research Europe, Zurich, S\"{a}umerstrasse 4,  CH-8803 R\"{u}schlikon, Switzerland}

\author{Johann Riemensberger}
\affiliation{Institute of Physics, Swiss Federal Institute of Technology Lausanne (EPFL), CH-1015 Lausanne, Switzerland}
\affiliation{Center of Quantum Science and Engineering (EPFL), CH-1015 Lausanne, Switzerland}
\affiliation{Present address: Department of Electronic Systems, Norwegian University of Science and Technology, 7491 Trondheim, Norway}

\author{Alisa Davydova}
\affiliation{Institute of Physics, Swiss Federal Institute of Technology Lausanne (EPFL), CH-1015 Lausanne, Switzerland}
\affiliation{Center of Quantum Science and Engineering (EPFL), CH-1015 Lausanne, Switzerland}

\author{Mikhail Churaev}
\affiliation{Institute of Physics, Swiss Federal Institute of Technology Lausanne (EPFL), CH-1015 Lausanne, Switzerland}
\affiliation{Center of Quantum Science and Engineering (EPFL), CH-1015 Lausanne, Switzerland}

\author{Paul Seidler}
\email[]{pfs@zurich.ibm.com}
\affiliation{IBM Research Europe, Zurich, S\"{a}umerstrasse 4,  CH-8803 R\"{u}schlikon, Switzerland}

\author{Tobias J. Kippenberg}
\email[]{tobias.kippenberg@epfl.ch}
\affiliation{Institute of Physics, Swiss Federal Institute of Technology Lausanne (EPFL), CH-1015 Lausanne, Switzerland}
\affiliation{Center of Quantum Science and Engineering (EPFL), CH-1015 Lausanne, Switzerland}
\affiliation{Institute of Electrical and Micro Engineering (IEM), Swiss Federal Institute of Technology Lausanne (EPFL), CH-1015 Lausanne, Switzerland}

\maketitle

\textbf{
Optical amplification, crucial for modern communication and data center interconnects, primarily relies on erbium-doped fiber amplifiers (EDFAs) to enhance signals without distortion~\cite{mears1987low,Desurvire:87}.
While EDFAs were historically decisive for the introduction of dense wavelength-division multiplexing, they only cover a portion of the low-loss spectrum of optical fibers.
This has motivated the development of amplifiers that operate beyond the erbium gain window, such as Raman amplifiers.
Pioneering work on optical traveling-wave parametric amplifiers (TWPAs)~\cite{Hansryd:02,marhic2015fiber} utilizing intrinsic third-order optical nonlinearity has led to demonstrations of increased channel capacity and performance.
TWPAs offer high gain, can reach the 3-dB quantum limit for phase-preserving amplifiers, and exhibit unidirectional operation.
Yet, despite the use of highly nonlinear fibers~\cite{marhic1996broadband,sudo1986high,Hansryd:02,Hansryd:01,torounidis2006fiber} or bulk waveguides~\cite{umeki2011phase,kishimoto2016highly}, their power requirements and technical complexity have impeded 
adoption.
In contrast, TWPAs based on integrated photonic circuits
offer the advantages of substantially increased mode confinement and optical nonlinearity but have been limited in bandwidth  
because of the trade-off with maintaining
low propagation loss~\cite{riemensberger2022photonic,ye2021overcoming}.
We overcome this challenge by using low-loss gallium phosphide-on-silicon dioxide~\cite{schneider2018gallium,wilson2020integrated,nardi2023soliton} photonic integrated circuits and attain up to 35~dB of parametric gain with waveguides only a few centimeters long in a compact footprint of only 0.25~square millimeters.
Fiber-to-fiber net gain exceeding 10~dB across an ultra-broad bandwidth of approximately 140~nm (\textit{i.e.}, 17~THz) is achieved, surpassing the gain window of a standard C-band EDFA by more than 3 times.
We furthermore demonstrate the capability to handle ultra-weak signals; input powers can range over six orders of magnitude while maintaining a low noise figure.
We exploit these performance characteristics to amplify both optical frequency combs and coherent communication signals.
This marks the first ultra-broadband, high-gain, continuous-wave amplification in a photonic integrated circuit, opening up new capabilities for next-generation interconnects in data centers, artificial-intelligence accelerators, and high-performance computing, as well as optical communication, metrology, and sensing.}\\

\noindent \textbf{Introduction}
Optical fibers and the broadband amplification of traveling- and continuous-wave (CW) optical signals have provided pivotal advancements in modern science and technology, particularly in the domain of optical communications for the transmission of information over long distances.
A crucial breakthrough for intercontinental optical fiber networks was the introduction of erbium-doped fiber amplifiers (EDFAs)~\cite{mears1987low,Desurvire:87} that could simultaneously amplify multiple wavelengths of light and eliminate the need for frequent electronic signal regeneration. EDFAs hence played a decisive role in the expansion of the World Wide Web.
Efforts have been made to extend the bandwidth of the optical gain by using Raman-assisted EDFAs~\cite{singh2012flat} and semiconductor optical amplifiers (SOAs)~\cite{connelly2007semiconductor}.
In contrast to amplifiers based on rare-earth-doped fibers or the more recently developed rare-earth-doped photonic integrated circuits (PICs)~\cite{liu2022photonic,gaafar2023femtosecond}, optical parametric amplifiers (OPAs) rely on the intrinsic nonlinearities of optical materials to generate gain.
Such traveling-wave parametric amplifiers (TWPAs) were originally developed in the microwave domain, where they, e.g., provide quantum-limited amplification for qubit readout~\cite{macklin2015near}.

OPAs have unique properties that set them apart from other means of amplification~\cite{Hansryd:02,marhic2015fiber}.
The shape of the amplification window is entirely determined by the optical dispersion and is limited only by the transparency of the material and its nonlinear absorption threshold.
The central frequency is adjustable, offering flexibility in signal processing.
Parametric amplification provides the capability of frequency conversion through the creation of a concomitant idler wave carrying the same information as the input signal but at a different frequency~\cite{tkach1995four,yoo1996wavelength}.
The reversal of the optical phase of the idler relative to that of the signal can be utilized to compensate for dispersion and mitigate nonlinear effects in transmission systems~\cite{ellis2016impact, pepper1980compensation,fisher1983optical}.
Kerr nonlinearity provides a nearly instantaneous response, allowing rapid operation of the OPA.
In contrast to the gain from erbium, parametric gain can be adjusted in situ by varying the optical drive power while maintaining a low noise figure, a behavior essential for amplification of weak input signals.
Moreover, optical TWPAs operate near the quantum noise limit and offer the ability to perform noiseless phase-sensitive amplification and thus have the potential to increase the span length of long-haul fiber-optic links~\cite{Tong:11,andrekson2020fiber}.
Lastly, OPAs are unidirectional, making them resistant to optical feedback and parasitic lasing, reducing the need for optical isolators in the optical path after the parametric amplifier, particularly in systems reliant on cascaded amplification chains.
The increased insensitivity to reflections from the chip facet suggests significant suppression of parasitic lasing within the OPA system, which can readily occur with high on-chip gain.
Although it is currently not feasible to remove the isolator from the EDFAs typically used to achieve high pump powers in OPA systems, they could be replaced by semiconductor laser diodes in the future.

To date, time-continuous operation of an OPA with substantial gain has only been demonstrated in fiber-based or bulk-crystal systems.
For example, OPAs can be constructed by employing hundreds of meters of highly nonlinear fibers~\cite{marhic1996broadband,sudo1986high,Hansryd:02,Hansryd:01,torounidis2006fiber}, but they require mitigation of strong detrimental Brillouin scattering.
Alternatively, mechanically cut lithium niobate waveguides~\cite{umeki2011phase,kishimoto2016highly} can be used, for which the $\chi^{(2)}$ nonlinearity requires periodic poling to generate quasi-phase matching and the performance is limited by the available wafer size, as the mode area is large and bending losses are excessive. 
In contrast, photonic integrated circuits (PICs) and particularly integrated waveguides, which are equivalent to fiber OPAs, offer the possibility of a significantly increased parametric gain coefficient $ g = \sqrt{ (\gamma P_{\textrm{P}})^2 - (\kappa/2)^2 }$~\cite{Hansryd:02,stolen1982parametric}, which is determined by the effective nonlinearity $ \gamma = {n_2 \omega_{\textrm{P}}}/{c A_\mathrm{eff}} $ and the phase-mismatch parameter $\kappa$; here, $ n_2 $ is the Kerr nonlinearity, $ A_\mathrm{eff} $ is the effective mode area, $ \omega_{\textrm{P}} $ is the pump angular frequency, and $ P_{\textrm{P}} $ is the pump power.
With the proper choice of material, PICs can exhibit a strong Kerr nonlinearity while simultaneously featuring a high linear refractive index that substantially reduces the mode area. 
The material should also have negligible two-photon absorption and a high optical-damage threshold. 
Integrated photonic platforms that have been studied so far include silicon~\cite{Foster:06,liu2010mid,Kuyken:11,gajda2012highly,Wang:15_SiH}, chalcogenides~\cite{Lamont:08}, Si$_7$N$_3$~\cite{Ooi:17}, AlGaAs~\cite{Pu:18}, and highly doped silica~\cite{ferrera2008low}.
None of these materials have yet produced net gain in the continuous pumping regime, as required for time-varying signals used in most applications.
Recently, triggered by advances in ultra-low-loss \SiN\ integrated photonic waveguides~\cite{liu2021high}, it was possible to achieve CW parametric amplification with net gain~\cite{ye2021overcoming,riemensberger2022photonic}.
However, this approach still requires coping with high pump-power levels, meter-long waveguides, limited gain, and narrow bandwidth.
Here, we address these challenges by demonstrating a compact, ultra-broadband, high-gain, PIC-based optical TWPA.\\

\noindent \textbf{Optical gain measurements}
We use thin-film gallium phosphide (GaP) on silicon dioxide to create a TWPA comprising a dispersion engineered waveguide operating at a pump wavelength near 1550~nm (see Supplementary Material for details of the fabrication).
\begin{figure*}[hbt!]
	\centering
	\includegraphics[width=1\textwidth]{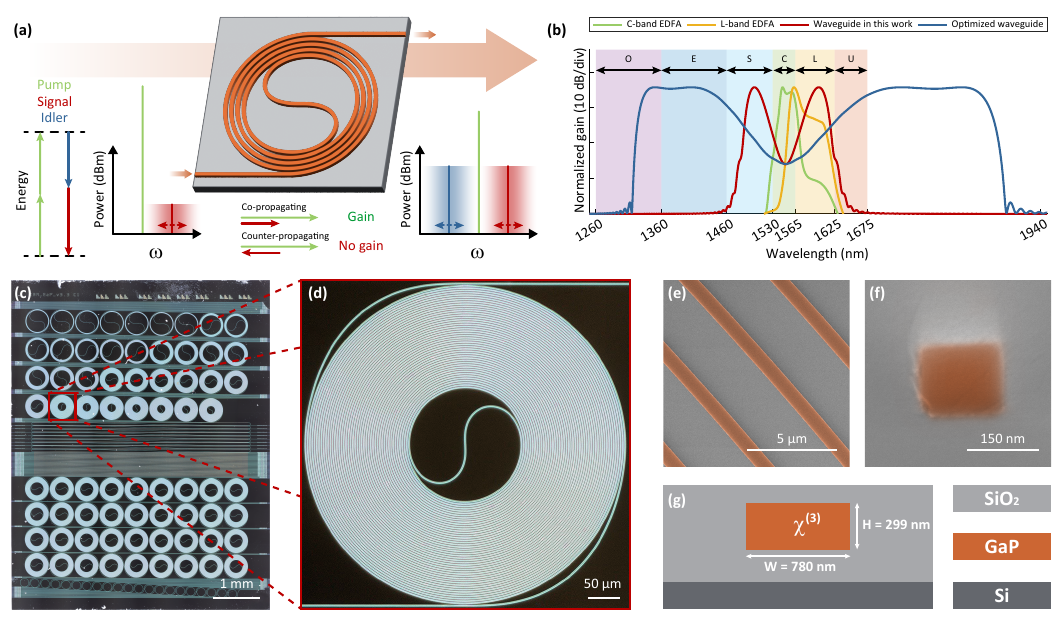}
	\caption{\textbf{Optical continuous-wave parametric amplification in a integrated GaP photonic waveguide.} 
		\textbf{(a)}~Principle of broadband optical parametric amplification in an integrated spiral waveguide. Strong pump and small signal waves co-propagate through the waveguide, leading to the amplification of the signal and generation of the third wave, the idler.
		The frequency of the signal wave can be chosen over a wide range within the amplification bandwidth.
		\textbf{(b)}~Comparison of the typical amplification bandwidth of state-of-the-art EDFAs (green: C-band, yellow: L-band) with the amplification bandwidth of integrated GaP parametric amplifiers (red: achieved in this work, blue: if further optimized).
		Gain values are provided for reference only, and not meant for direct comparison.
		\textbf{(c)}~Optical microscope image of the fabricated photonic chip with multiple spiral waveguides and other test structures.
		\textbf{(d)}~Enlarged optical microscope image of a 5-cm-long spiral waveguide. Accounting for the chip width of 0.55~cm, the total waveguide length is 5.55~cm.
		\textbf{(e)}~Scanning electron microscope image of a few waveguide turns.
		\textbf{(f)}~Scanning electron microscope image of an inverse taper facet, similar to the one used in this work, used to couple light into a GaP waveguide. 
		\textbf{(g)}~Material stack of the fabricated chip.}
	\label{fig:gap_opa}
\end{figure*}
GaP combines a high optical refractive index $(n = 3.05)$ with a strong Kerr nonlinearity ($n_2 = 1.1 \times 10^{-17}$~m$^2$~W$^{-1}$) and an indirect bandgap~\cite{lide2004crc} ($E_\mathrm{g} = 2.24$~eV), sufficiently large to mitigate two-photon absorption at telecommunication wavelengths. This exceptional confluence of properties has facilitated the generation of low-threshold frequency combs~\cite{wilson2020integrated} and dissipative Kerr solitons~\cite{nardi2023soliton}.
We estimate the effective nonlinearity of the GaP waveguides in this work to be $\gamma = 165$~W$^{-1}$m$^{-1}$, which is more than 300 times larger than the value of 0.51~W$^{-1}$m$^{-1}$ reported for \SiN\ waveguides~\cite{riemensberger2022photonic}, facilitating a 35-fold reduction in waveguide length and a 60-fold reduction in device footprint. 
Advances in fabrication techniques~\cite{wilson2020integrated,nardi2023soliton} allow us to reduce GaP waveguide optical propagation losses to 0.8~dB~cm$^{-1}$ on average within the S-, C- and L-band (see Supplementary Material), and achieve a high bonding yield and low defect density. 
These developments are pivotal for successful fabrication of centimeter-long waveguide spirals with a footprint of only 500~$\times$~500~$\mu$m$^2$ (Figs.~\ref{fig:gap_opa}(c~-~g) and Supplementary Material).
\begin{figure*}[hbt!]
	\centering
	\includegraphics[width=1\textwidth]{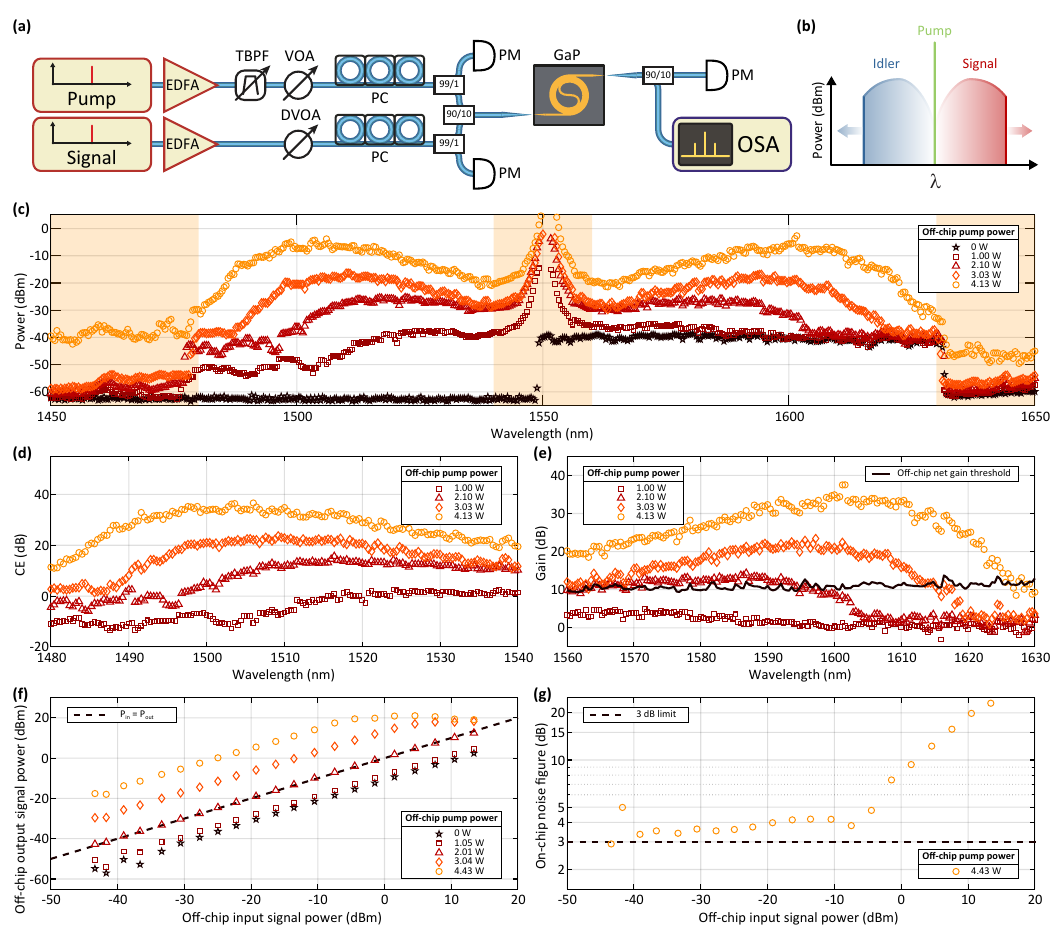}
	\caption{\textbf{Observation of a broadband continuous-wave parametric amplification in a GaP spiral waveguide.}
			\textbf{(a)}~Schematic of the experimental setup.
			EDFA: erbium-doped fiber amplifier; (D)VOA: (digital) variable optical attenuator; PC: polarization controller; PM: power meter; OSA: optical spectrum analyzer.
			\textbf{(b)}~Principle of the "Max Hold" setting of the OSA used in broadband gain measurements. With every new scan, the OSA records and updates only the highest values across the measurement span, while the wavelength of the signal laser is slowly swept to cover the whole amplification bandwidth.
			\textbf{(c)}~Measured amplification spectra for various off-chip pump powers.
			The resolution bandwidth of the OSA is 2~nm.
			EDFA and DVOA in the signal path are not used in this experiment.
			\textbf{(d,~e)}~Calculated idler conversion efficiency (CE) and signal amplification spectra for various input pump powers up to 4.13~W.
			\textbf{(f)}~Off-chip signal power at wavelength 1605~nm at the output of the amplifier as a function of signal and pump powers at the input fiber.
			The resolution bandwidth of the OSA is 0.1~nm.
			The dashed line is the boundary at which the input signal power is equal to the output signal power.
			\textbf{(g)}~Measurement of the optical TWPA noise figure as a function of signal input power.
			The dashed line indicates the 3~dB quantum limit of a phase-insensitive amplifier.}
	\label{fig:gap_opa_gain}
\end{figure*}
We operate the TWPA with a single pump to amplify signals via degenerate four-wave-mixing (DFWM), exploiting the optical Kerr effect (Fig.~\ref{fig:gap_opa}(a)).
The optical energy is redistributed by annihilating two pump photons to amplify signal photons at a separate frequency, while simultaneously generating a phase-conjugated idler at a frequency equidistant from the pump laser but offset in the opposite direction.
The near instantaneous action of the optical Kerr effect necessitates phase-matching in the waveguide for efficient amplification~\cite{Hansryd:02}, i.e., $ \kappa = \Delta \beta + 2 \gamma P_{\textrm{P}} $ is small, where $ \Delta \beta $ is the linear propagation mismatch given by $ \Delta \beta \approx \beta_2\left(\omega_\textrm{S}-\omega_\textrm{P}\right)^2 + 	\beta_4\left(\omega_\textrm{S}-\omega_\textrm{P}\right)^4/12 + ... $, $\beta_2$ and $\beta_4$ are the second- and fourth-order derivatives with respect to the angular frequency $\omega$ of the optical propagation constant $\beta$ evaluated at the pump frequency $ \omega_{\textrm{P}} $, and $ \omega_{\textrm{S}} $ is the signal angular frequency.
Only even order dispersion terms contribute to the shape of the amplification spectrum.
The largest optical bandwidth is achieved when the dispersion is slightly anomalous ($\beta_2 < 0$) and the linear phase mismatch $\Delta \beta$ is compensated by the nonlinear phase mismatch $ 2 \gamma P_{\textrm{P}} $ originating from self- and cross-phase modulation.
The peak gain in a waveguide of length $ L $ is defined by $ G_{\textrm{S}} = 1 + \left[ \sinh(-\Delta \beta L_{\textrm{eff}}/2) \right]^2 $, where $ L_{\textrm{eff}} = (1-\exp(-\alpha L))/\alpha $ is the effective length and $ \alpha $ is the linear propagation loss.
Fig.~\ref{fig:gap_opa}(b) shows a comparison of the gain spectrum calculated for an optimized GaP optical TWPA with that of C-band and L-band EDFAs.
The optimized GaP waveguide has a cross-section of $ 789\times299$~$\text{nm}^2 $, for which $\beta_2 =-16$~fs$^2$mm$^{-1}$ and $\beta_4 = 3547$~fs$^4$mm$^{-1}$. The region of signal and idler gain extends from around 1300~nm in the optical O-band to 1900~nm, well beyond the longest wavelengths used in silica fiber communications.
We find that the waveguides used in this work exhibit a stronger anomalous dispersion, $\beta_2 = -124$~fs$^2$mm$^{-1}$, than expected from our simulations (see Supplementary Material). 
We measure the CW amplification spectrum of a 5.55~cm-long GaP spiral waveguide with a height of 299~nm and a design width of 780~nm using the same methods as described in~Ref.~\cite{riemensberger2022photonic}.
Figures.~\ref{fig:gap_opa_gain}(a,~b) depict the experimental setup and recording scheme used to measure the parametric gain (see Supplementary Material), and 
Fig.~\ref{fig:gap_opa_gain}(c) shows the power-calibrated amplification spectrum.
In the spectral region around the pump wavelength, we observe a broadened peak due to the leakage of the amplified spontaneous emission from the EDFA installed in the optical path of the pump; our tunable filter has a bandwidth of 1~nm and a finite suppression ratio of 30~dB. 
The frequency conversion and amplification spectra (Figs.~\ref{fig:gap_opa_gain}(d,~e)) are determined by comparing the output spectra recorded as the signal laser (set to 1.5~$\mu$W) is swept at various pump power levels (red to orange) with the output spectrum obtained when the pump is switched off (black).
Increasing the input power from 1.00~W to 4.13~W, we observe increasing gain, conversion efficiency, and amplification bandwidth, following established theory~\cite{Stolen:82,Hansryd:02}.
The exponential increase in gain with a linear change in pump power is observed, as expected.
The generated idler spectrum spans from 1550~nm to approximately 1480~nm at the highest pump power.
In Fig.~\ref{fig:gap_opa_gain}(c), spurious light generated outside the scan range of the signal laser (outer shaded regions) originates from non-degenerate FWM. 
Hence, we only calculate the internal idler conversion efficiency (Fig.~\ref{fig:gap_opa_gain}(d)) from 1480~nm to 1540~nm and the internal signal gain from 1560~nm to 1630~nm (Fig.~\ref{fig:gap_opa_gain}(e)). 
We separately measure the passive transmission loss of the TWPA, i.e., the insertion loss, by bypassing the lensed fibers and the chip; the ratio of the spectrum obtained with the bypass to the spectrum transmitted through the chip is plotted as a black line in Fig.~\ref{fig:gap_opa_gain}(e).
To calculate the off-chip net gain, we subtract this 10 -- 12~dB of insertion loss, revealing net-gain of up to $G_\textrm{S} = 25$~dB, with $G_\textrm{S} > 10$~dB over a bandwidth of 70~nm. 
Assuming a symmetric gain spectrum on the short-wavelength side of the pump, we estimate the 10~dB off-chip gain region to exceed 140~nm, or 17~THz.
This bandwidth is almost ten times larger than reported in prior work on \SiN~\cite{riemensberger2022photonic}, and nearly three times larger than the bandwidth of a single C-band EDFA.
To investigate the maximum signal output power and the saturation power of the amplifier, we perform a power sweep of the signal at a wavelength of 1605~nm (Fig.~\ref{fig:gap_opa_gain}(f)).
Here we use a different spiral with a designed waveguide width of 790~nm; this spiral is used also in the remaining experiments.
The measurements are carried out with the resolution bandwidth of the OSA set at 0.1~nm. 
At a pump power of 4.43~W, the saturated off-chip signal output power exceeds 125~mW, corresponding to a power of 220~mW on chip at the waveguide output.
Hence, the on-chip power conversion efficiency reaches 9~\%, which is limited by the waveguide propagation losses.
Here, we also demonstrate the ability to achieve low-noise amplification with high gain over a large dynamic range.
The linear, small-signal regime of amplification extends up to a signal input power of almost 0~dBm at the highest pump power level of 4.43~W.
\begin{figure*}[hbt!]
	\centering
	\includegraphics[width=1\textwidth]{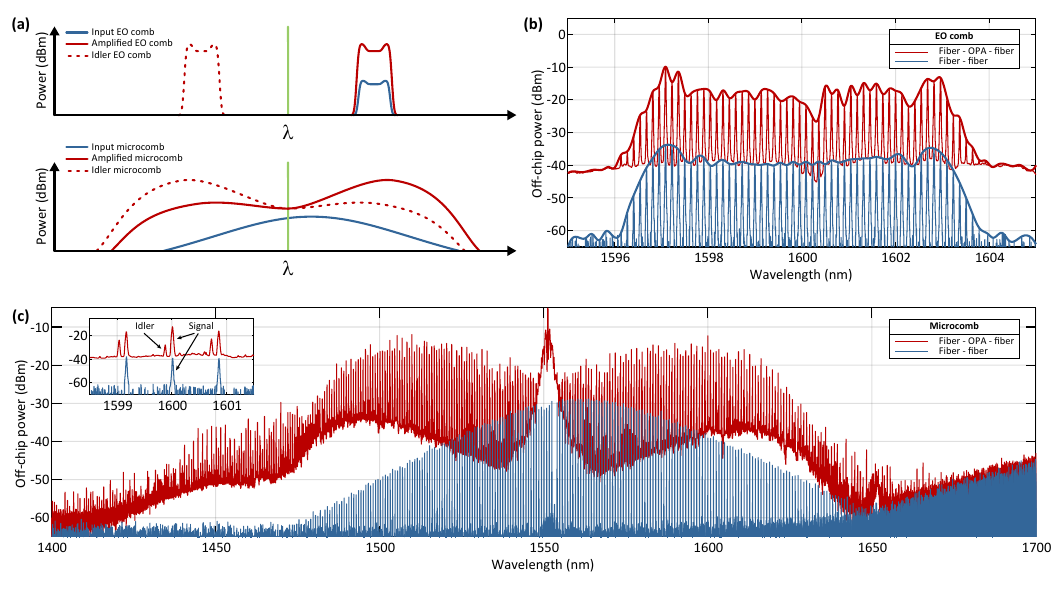}
	\caption{\textbf{Amplification of optical frequency combs using a GaP optical TWPA. } 
		\textbf{(a)}~Schematic of amplification and frequency conversion of a EO frequency comb (top) and a dissipative Kerr soliton frequency comb (bottom).
		\textbf{(b)}~Spectra of the amplified (red) and input (blue) EO frequency comb showing more than 20 dB fiber-to-fiber net gain.
		Envelopes are added for better visibility.
		\textbf{(c)}~Same as \textbf{(b)} but for amplification of a 100-GHz frequency comb produced by a single soliton state in a \SiN\ microring cavity.
		The inset shows the appearance of frequency-converted idler-comb lines as a result of DFWM.
		The increased noise floor at longer wavelengths in both spectra is due to the proximity to the end of the OSA measurement range.
		The resolution bandwidth of the OSA is 0.02~nm in both experiments.}
	\label{fig:gap_opa_combs}
\end{figure*}
At 2.01~W pump power, the amplification is linear over the entire range of input powers measured, extending over six orders of magnitude and limited only by the capabilities of our measurement setup; in principle, parametric amplification poses no lower limit on the signal input power.
We calculate the amplifier noise figure from the measured signal amplification and the spontaneous parametric fluorescence background (see Supplementary Material)~\cite{baney2000theory}.
The off-chip noise figure approaches 6~dB in the small-signal gain regime.
Since degradation of the signal-to-noise ratio (SNR) at the input facet is irreversible, we take into account the coupling losses of 2.5~dB at the chip input facet and find the on-chip noise figure of less than 4~dB for a wide range of signal powers below saturation (Fig.~\ref{fig:gap_opa_gain}(g)).
The degradation of the noise figure at higher signal input powers is attributed to gain saturation; in other words, the signal gain becomes weaker while the parametric fluorescence is still within the small-signal regime and undergoes a higher amplification.

\noindent \textbf{Amplification of optical frequency combs}
To highlight the application potential of such a broadband optical TWPA, we perform additional amplification experiments using low-power and high-repetition-rate frequency comb sources as signals.
In Fig.~\ref{fig:gap_opa_combs}(a), we schematically show amplification and idler generation for two types of injected frequency combs, a narrowband electro-optic (EO) comb~\cite{nardi2023soliton} and a broadband dissipative Kerr soliton comb~\cite{kippenberg2018dissipative}.
The EO comb is centered around 1600~nm and has a 16-GHz line spacing and a total signal power of 10.5~$\mu$W.
\begin{figure*}[htb!]
	\centering
	\includegraphics[width=1\textwidth]{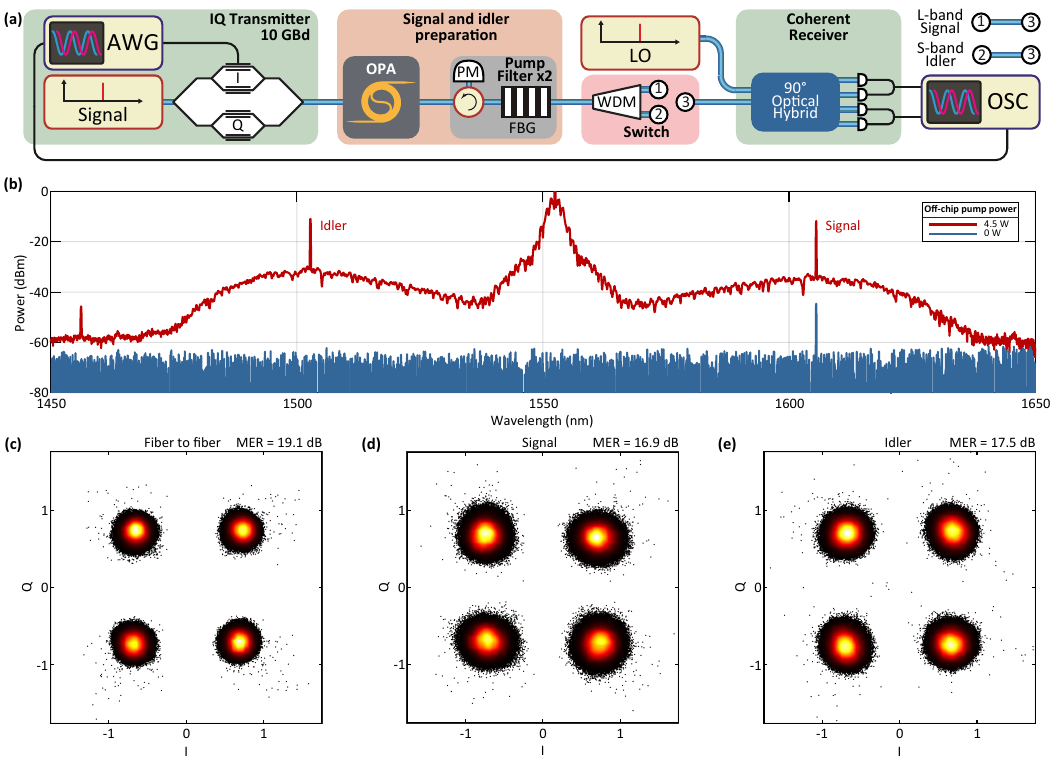}
	\caption{\textbf{Coherent data transmission experiment.} 
		\textbf{(a)}~Simplified schematic of the experimental setup used in the communication experiment. AWG: arbitrary waveform generator; FBG: fiber Bragg grating; WDM: wavelength-division multiplexer; OSC: oscilloscope; LO: local oscillator.
		\textbf{(b)}~Calibrated optical spectra measured at the output of the chip using the same OSA as described above (not shown here in \textbf{(a)}).
		The resolution bandwidth is set to 0.2~nm for faster acquisition.
		\textbf{(c,~d,~e)}~Constellation diagram and MER for three measurements: fiber-to-fiber reference, amplified signal, and generated idler.}
	\label{fig:opa_comm}
\end{figure*}
When transmitted through the amplifier together with a 4.12-W pump at 1550~nm, we observe more than 20~dB fiber-to-fiber net gain (Fig.~\ref{fig:gap_opa_combs}(b)).
The reference dataset (blue) is measured by bypassing the lensed fibers and the photonic chip.
Note that, due to the nearly instantaneous nature of parametric amplification, we do not use a dispersive fiber compression stage during the EO comb preparation before the amplifier to avoid high signal peak powers and amplifier saturation.
We perform the same experiment with the frequency comb formed by a dissipative Kerr soliton state~\cite{kippenberg2018dissipative} generated in a \SiN\ microring resonator.
The total power of the soliton comb is 76~$\mu$W, and the repetition rate is 100~GHz.
In the spectral region of the highest gain, far from the soliton pump located at 1552.8~nm, the input power of individual lines decreases to the level of a few nW, while lines with the highest initial power, located closer to the pump, experience smaller gain. 
As expected, an idler comb mirrored with respect to the pump frequency is generated, and the line density of the output spectrum is doubled (Fig.~\ref{fig:gap_opa_combs}(c),~red).
Additional lines can be observed throughout the entire spectrum and, most notably, closer to the pump; we attribute their appearance to numerous FWM processes between various interacting lines.
Additional lines that are located at frequencies far outside the bandwidth of the input comb arise due to cascaded non-degenerate FWM processes.
A small feature in the amplified spectra around 1651~nm is observed with all available waveguides and can be attributed to Raman scattering in GaP~\cite{wilson2020integrated,hobden1964raman}.
The flattened spectra resulting from amplification provide clear evidence that the GaP OPA can handle the simultaneous input of a few hundred lines over a broad bandwidth, and the parametric gain remains as high as for a single input frequency if operated in the small-signal regime.
Moreover, we again observe amplification of very small input signals, with the lowest comb lines being as weak as 1~nW, something that is not possible with typical EDFAs. 
The amplification of such small signals may be especially attractive for applications such as LiDAR~\cite{riemensberger2020massively} or free-space communications.

\noindent \textbf{Coherent communication and modulation transfer}
To further demonstrate the potential of the GaP OPA for use in coherent communication, we set up a 10~GBd communication line, as shown in Fig.~\ref{fig:opa_comm}(a).
We send a QPSK-encoded pseudo-random bit sequence at a signal wavelength of 1605~nm (L-band) to our OPA and then perform heterodyne measurements of the amplified signal.
Additionally, using another laser as a local oscillator (LO), we measure modulation transfer to the idler located at approximately 1502~nm (S-band).
The signal wavelength is chosen to be in the region of maximum gain, while also considering the limitations of the available equipment.
Fig.~\ref{fig:opa_comm}(b) shows the corrected OPA output spectra with the pump switched on and off.
The internal gain achieved exceeds 30~dB, which is enough to compensate for coupling and propagation losses on the chip (which barely exceed 10~dB in total at the selected signal wavelength) and even for 10~dB of losses on the splitter before the chip and 4~dB of losses in the filtering section and wavelength-division multiplexer, so we achieve a positive net gain accounting for all losses present in the system (see Supplementary Material).
The received signal power is 26~$\mu$W and the idler power is 32~$\mu$W; the idler power is higher than the signal power due to lower losses at the idler wavelength.
The LO power is 29.1~mW for the signal measurement and 27.7~mW for the idler.
As a reference, we measure the signal directly sent from the transmitter to the receiver through a fiber, bypassing all the other components without attenuation.
We digitally post-process and analyze the data and plot constellation diagrams for all three measurements -- for the fiber reference, signal, and idler (Figs.~\ref{fig:opa_comm}(c,~d,~e), respectively).
As a figure of merit, for each measurement we estimate the modulation error ratio (MER)~\cite{ETR290}, defined as the sum of the squares of the magnitude of the ideal symbol vectors divided by the sum of the squares of the magnitude of the symbol error vectors (see Supplementary Material).
The idler conversion efficiency is high enough in the GaP OPA to generate an idler of the same power level as that of the amplified signal.
This eliminates any need for additional idler post-amplification, as it is sufficiently strong to be used directly for inter-band signal translation.\\

\noindent \textbf{Discussion}
In summary, using an integrated GaP photonic platform, we achieve both high net gain and broadband phase-insensitive operation of an OPA.
The maximum fiber-to-fiber net gain reaches 25~dB, and the combined signal and idler 10-dB-gain bandwidth is 140~nm, substantially greater than the bandwidth of both EDFAs and existing CW parametric amplification systems~\cite{kishimoto2016highly,torounidis2006fiber,riemensberger2022photonic}.
We demonstrate the potential for practical application of GaP waveguides for the amplification of frequency combs, coherent optical data streams, and optical signals over more than 60~dB of dynamic range, fully covering and extending beyond input powers from -26~dBm to 0~dBm, and output powers from -8~dBm to 6~dBm, which are typical requirements in telecommunication technology~\cite{ITU-T_G.698.2}.
The ability to amplify weak optical signals over a large dynamic range may be decisive for a variety of applications beyond coherent communications, including LiDAR and sensing.
Moreover, integrated TWPA can be utilized where amplifiers are needed in custom bands, such as 1650~nm for methane detection or in applications like optical coherence tomography requiring amplification at 1300~nm, where no chip-scale solutions based on rare-earth ions currently exist.

Further reducing the optical propagation losses will greatly lower the required pump power and may even allow direct pumping of the OPA with semiconductor lasers.

Our on-chip GaP OPA solves some of the most pivotal challenges that have prevented the widespread adoption of fiber OPA systems.
Viewed more broadly, the results mark an example where the performance of nonlinear photonic integrated circuit-based amplifiers surpasses that of legacy fiber-based systems.
The dispersion of the waveguide can be designed as needed because it is defined lithographically, and the waveguide's short length ensures a broad bandwidth~\cite{ayan2023towards} and reduces sensitivity to fabrication imperfections, in contrast to optical fibers that are hundreds of meters long.
Compared to other photonic-chip-based TWPAs~\cite{riemensberger2022photonic,ye2021overcoming}, 
the GaP TWPA features a waveguide length and overall device footprint that are reduced by orders of magnitude.
The low Brillouin gain of integrated waveguides eliminates the technical complexity of adding pump phase modulators~\cite{Hansryd:01,torounidis2006fiber,Tong:11,Gyger:20}.
In principle, the pump frequency can be freely chosen over a wide frequency range when designing the TWPA, as it is not restricted by an excitation spectrum as in doped amplifiers. 
Also, in contrast to existing EDFA technology, the optical TWPA is unidirectional, making it more robust against parasitic reflections, which not only remain unamplified but are actually further suppressed due to the propagation losses.
Lastly, the TWPA can in principle achieve a lower noise figure than an EDFA and lower than what we have demonstrated here, as it can be operated as a phase-sensitive amplifier~\cite{andrekson2020fiber,ye2021overcoming}, reducing the noise below the quantum limit and simultaneously increasing the peak gain by 6~dB.

Our results signal the emergence of compact, high-performance photonic integrated circuit based optical integrated TWPAs with large bandwidth, high gain and small footprint that have the potential to transition from the laboratory into future optical communication systems.
Broadband optical TWPAs may find use in a variety of  future system configurations.
For example, nonlinear distortions in parametrically amplified fiber links may be mitigated by alternately propagating the signal and phase-conjugated idler through successive fiber link sections~\cite{solis2015optimized}.

\footnotesize
\noindent \textbf{Author contributions:}
N.K. and J.R. performed numerical simulations of dispersion and parametric gain, and designed spiral waveguides.
A.N. fabricated the sample.
J.R. and N.K. performed the initial linear characterization of the devices.
N.K. and A.N. carried out the broadband gain measurements.
N.K. and A.D. conducted frequency comb amplification experiments.
N.K. and M.C. executed the communication experiment.
N.K. carried out the rest of the measurements presented in this work.
N.K. and J.R. prepared the manuscript with contributions from all authors. 
J.R., P.S. and T.J.K supervised the work.

\noindent \textbf{Funding information:}
This work was supported by the European Union's Horizon 2020 research and innovation programme under the Marie Sk{\l}odowska-Curie grant agreement No 812818 (MICROCOMB), by the Swiss National Science Foundation (SNF) under grant number 216493 (HEROIC), and by the Air Force Office of Scientific Research under award number FA9550-19-1-0250.

\noindent \textbf{Acknowledgments:}
The sample was fabricated at the Binnig and Rohrer Nanotechnology Center (BRNC) at IBM Research Europe, Zurich. 

\noindent \textbf{Disclosures:}
J.R., N.K. and T.J.K. are inventors on European Patent Application No. 24165690.9 submitted by Swiss Federal Institute of Technology Lausanne (EPFL) that covers "Optical parametric amplifier apparatus and optical signal transmission apparatus, and applications thereof".
Other authors declare no competing interests.

\noindent \textbf{Data availability:}
All experimental datasets and scripts used to produce the plots in this work will be uploaded to the Zenodo repository upon publication of this preprint.

\bibliography{bibliography}

\end{document}


\title{Supplementary Material to:\\An ultra-broadband photonic-chip-based\\traveling-wave parametric amplifier}

\author{Nikolai Kuznetsov}
\thanks{These authors contributed equally to this work.}
\affiliation{Institute of Physics, Swiss Federal Institute of Technology Lausanne (EPFL), CH-1015 Lausanne, Switzerland}
\affiliation{Center of Quantum Science and Engineering (EPFL), CH-1015 Lausanne, Switzerland}

\author{Alberto Nardi}
\thanks{These authors contributed equally to this work.}
\affiliation{Institute of Physics, Swiss Federal Institute of Technology Lausanne (EPFL), CH-1015 Lausanne, Switzerland}
\affiliation{IBM Research Europe, Zurich, S\"{a}umerstrasse 4,  CH-8803 R\"{u}schlikon, Switzerland}

\author{Johann Riemensberger}
\affiliation{Institute of Physics, Swiss Federal Institute of Technology Lausanne (EPFL), CH-1015 Lausanne, Switzerland}
\affiliation{Center of Quantum Science and Engineering (EPFL), CH-1015 Lausanne, Switzerland}
\affiliation{Present address: Department of Electronic Systems, Norwegian University of Science and Technology, 7491 Trondheim, Norway}

\author{Alisa Davydova}
\affiliation{Institute of Physics, Swiss Federal Institute of Technology Lausanne (EPFL), CH-1015 Lausanne, Switzerland}
\affiliation{Center of Quantum Science and Engineering (EPFL), CH-1015 Lausanne, Switzerland}

\author{Mikhail Churaev}
\affiliation{Institute of Physics, Swiss Federal Institute of Technology Lausanne (EPFL), CH-1015 Lausanne, Switzerland}
\affiliation{Center of Quantum Science and Engineering (EPFL), CH-1015 Lausanne, Switzerland}

\author{Paul Seidler}
\email[]{pfs@zurich.ibm.com}
\affiliation{IBM Research Europe, Zurich, S\"{a}umerstrasse 4,  CH-8803 R\"{u}schlikon, Switzerland}

\author{Tobias J. Kippenberg}
\email[]{tobias.kippenberg@epfl.ch}
\affiliation{Institute of Physics, Swiss Federal Institute of Technology Lausanne (EPFL), CH-1015 Lausanne, Switzerland}
\affiliation{Center of Quantum Science and Engineering (EPFL), CH-1015 Lausanne, Switzerland}
\affiliation{Institute of Electrical and Micro Engineering (IEM), Swiss Federal Institute of Technology Lausanne (EPFL), CH-1015 Lausanne, Switzerland}

\maketitle

{\hypersetup{linkcolor=black}\tableofcontents}

\newpage

\renewcommand{\thefigure}{S\arabic{figure}}
\renewcommand{\theequation}{S\arabic{equation}}

\section{Device fabrication}

We fabricate the GaP photonic chip using a process similar to that outlined in the Supplementary Information of Ref.~\cite{wilson2020integrated}.
We epitaxially grow thin films by metal-organic chemical vapor deposition on a sacrificial [100]-oriented GaP wafer that is re-polished before growth to mitigate the formation of hillocks. Hillocks form due to surface contamination of the as-received commercial GaP wafers and lead to bonding defects. 
The grown layers consist of a 100-nm-thick GaP buffer layer, a 200-nm-thick Al$_{0.2}$Ga$_{0.8}$P layer (later used as an etch-stop layer to selectively remove the sacrificial wafer), and a 299-nm-thick GaP device layer. 
The surfaces of both the GaP wafer following growth and of a silicon wafer capped with 2 $\mu$m of SiO$_2$ are prepared for wafer bonding by depositing 5~nm of Al$_2$O$_3$ by thermal atomic-layer deposition. 
After bonding, the wafer is annealed to improve the strength and stability of the bond. 
The sacrificial GaP wafer is then removed by mechanical grinding, down to a thickness of about 50~$\mu$m. 
The remaining portion of the sacrificial GaP wafer is dry etched on chip level in a mixture of SF$_6$ and SiCl$_4$ in an inductively coupled-plasma reactive-ion-etching (ICP-RIE) process~\cite{honl2018highly}.
The etch rate slows substantially once the Al$_{0.2}$Ga$_{0.8}$P layer is exposed to the plasma.
Subsequently, the Al$_{0.2}$Ga$_{0.8}$P stop layer is selectively removed by wet etching in concentrated HCl for 4 minutes.
The chip's surface is promptly covered with 3~nm of SiO$_2$ deposited by plasma atomic-layer deposition at a temperature of 300~$^\circ$C.
This thin layer of SiO$_2$ also acts as an adhesion promoter for the negative resist hydrogen silsesquioxane (HSQ) that is employed to pattern the devices via electron-beam lithography.
The spirals are designed to fit in a single 525$\times$525~$\mu$m electron-beam write field to mitigate possible scattering losses that may originate from the imperfect stitching of neighboring write fields.
The resist pattern is transferred into the GaP by ICP-RIE using a mixture of BCl$_3$, Cl$_2$, CH$_4$, and H$_2$~\cite{schneider2018gallium}, after which the HSQ is removed by dipping the chip in buffered HF for 10 s.
A 2-$\mu$m-thick SiO$_2$ top cladding is applied via plasma-enhanced chemical vapor deposition using tetraethyl orthosilicate (TEOS) as precursor at 400~$^\circ$C.
Efficient input and output coupling of the light is enabled by 250-$\mu$m-long inverse tapers with a design tip width of 180~nm.
The edges of the chip are removed by subsurface-absorption laser dicing to expose the inverse tapers, providing clean vertical chip facets and efficient fiber-to-chip coupling to reduce the overall optical insertion loss.

\section{Dispersion engineering of GaP waveguides for optical parametric amplification}

\begin{figure*}[b]
	\centering
	\includegraphics[width=1\textwidth]{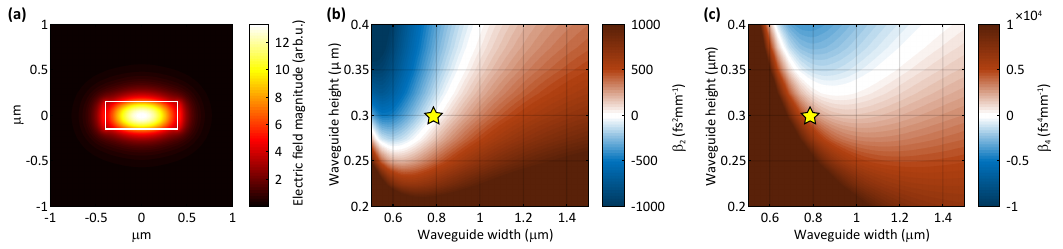}
	\caption{\textbf{Dispersion of integrated GaP waveguides.}
		\textbf{(a)} Electric field profile of the fundamental TE mode in a rectangular GaP waveguide with a width of 790~nm and a height of 299~nm.
		The mode is strongly confined due to the high refractive index contrast between the GaP core and the SiO$_2$ cladding.
		\textbf{(b, c)} Second- and fourth-order dispersion maps, respectively.
		The anomalous dispersion is available for waveguides with a height above 270~nm.
		Stars indicate the regions of optimal parameters necessary to achieve the broadband operation of the optical traveling-wave parametric amplifier made on a chip with a predefined GaP thickness of 299~nm.
		While slightly anomalous dispersion is typically required for parametric amplification because it compensates for nonlinear phase mismatch, the fourth-order dispersion term should be positive to counteract the excess propagation phase mismatch that accumulates due to second-order dispersion, thereby achieving broadband operation.}
	\label{fig:dispersion}
\end{figure*}
In strongly confining integrated waveguides, such as those made of GaP, it is possible to tune the group velocity dispersion $\beta_2$ and the zero-dispersion wavelengths over a wide range by varying the cross-sectional geometry~\cite{Turner:06,Okawachi:11}.
We, therefore, perform dispersion simulations of straight rectangular GaP waveguides fully cladded with SiO$_2$ for a range of film thicknesses and waveguide heights and widths using a commercially available finite-element-method solver, COMSOL Multiphysics\textsuperscript{\textregistered}, as depicted in Fig.~\ref{fig:dispersion}.
We find that the thickness of the GaP cannot be less than $\sim270$~nm; otherwise, dispersion is normal and only a relatively weak parametric gain is possible in a narrow window near the pump wavelength.

For the prepared GaP layer thickness of 299~nm, we design a waveguide with a width of 790~nm to operate in the anomalous dispersion regime.
We also fabricate waveguides with smaller widths, varied in steps of 5~nm.
Note that, because of the strong mode confinement, the difference between the dispersion of a straight waveguide and that of a bent waveguide is negligible for bending radii $\gtrsim$ 50~$\mu$m.
In our Archimedean spiral waveguides, the bending radius changes from 250~$\mu$m to approximately 80~$\mu$m. 
It may be possible to increase the amplification bandwidth further by increasing the GaP waveguide height to $\sim330$~nm because the contribution of the fourth-order dispersion parameter is near zero and varies less with the waveguide width at this thickness, thus leading to more broadband amplification and a more robust waveguide design. 

In our main experiments, we use waveguides that are 5.55 cm long, including the straight waveguide sections and the input and output inverse coupling tapers in addition to the spiral itself.
The parametric amplification bandwidth depends on the accumulated phase mismatch between pump, signal, and idler and, therefore, on the propagation constant mismatch and the device length~\cite{ayan2023towards}.
Hence, it is generally beneficial to use shorter devices to prevent bandwidth narrowing.
Moreover, short devices are less sensitive to cross-section variations that could potentially lead to additional bandwidth degradation.

\section{Numerical calculations of optical parametric gain}

\begin{figure*}[h!]
	\centering
	\includegraphics[width=1\textwidth]{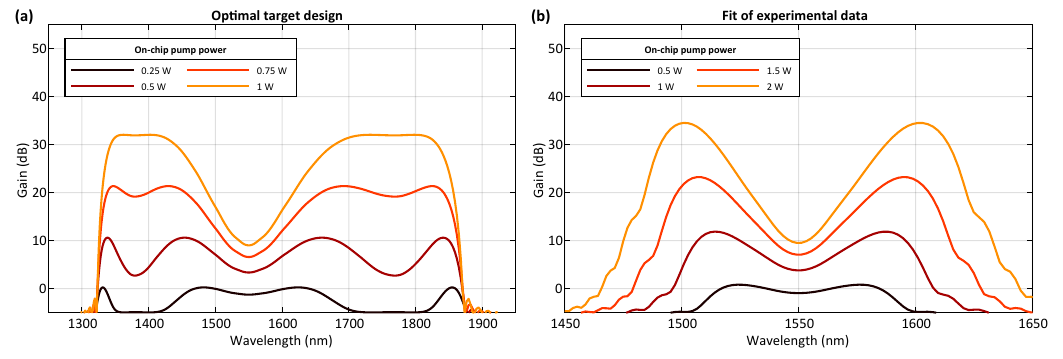}
	\caption{\textbf{OPA gain simulations.}
		\textbf{(a)} Gain in the waveguide with a theoretically optimal cross-section of  789$\times$299~$\text{nm}^2 $.
		At lower pump powers, two additional distinct gain peaks appear separate from the main part of the gain spectrum, and their positions correspond to the frequencies where the nonlinear phase-matching condition is satisfied with the help of the fourth-order dispersion term.
		These peaks merge with the central part as the pump power increases.
		\textbf{(b)} Gain in the waveguide with dispersion and nonlinearity tuned to match our experimental observations.
		Due to strong anomalous dispersion, the contribution of the fourth-order dispersion term diminishes, causing the two additional gain peaks to move well beyond the simulation wavelength window.
		Hence, the gain profile contains only two main gain lobes.}
	\label{fig:gain_simulations}
\end{figure*}

\begin{figure*}[h]
	\centering
	\includegraphics[width=1\textwidth]{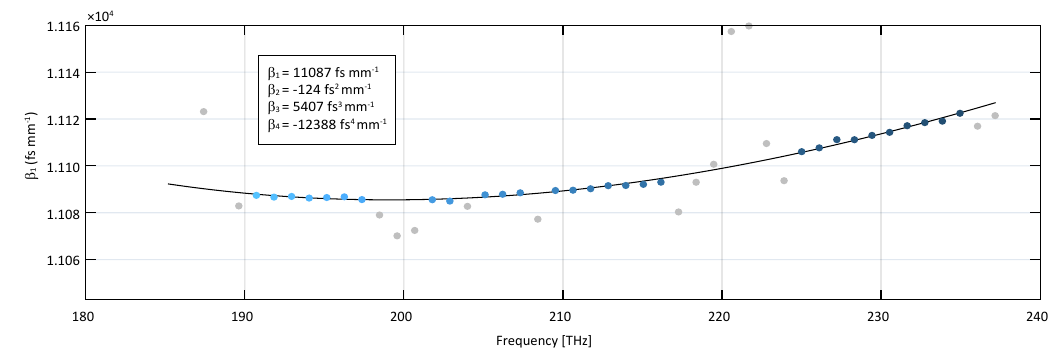}
	\caption{\textbf{Dispersion measurements of a 790-nm-wide spiral waveguide using frequency comb-assisted spectroscopy.}
		Dispersion measurements of a GaP waveguide are performed using OFDR by analyzing the delay between reflections from the input and output facets at different frequencies to determine the waveguide's dispersion characteristics.
		Grey dots are excluded from the dataset before fitting due to their large deviation.
		The values for $\beta_3$ and $\beta_4$ are given for reference and are not reliable due to the inaccuracy of the dispersion measurements of an extremely short waveguide.}
	\label{fig:dispersion_meas}
\end{figure*}

\begin{figure*}[h]
	\centering
	\includegraphics[width=1\textwidth]{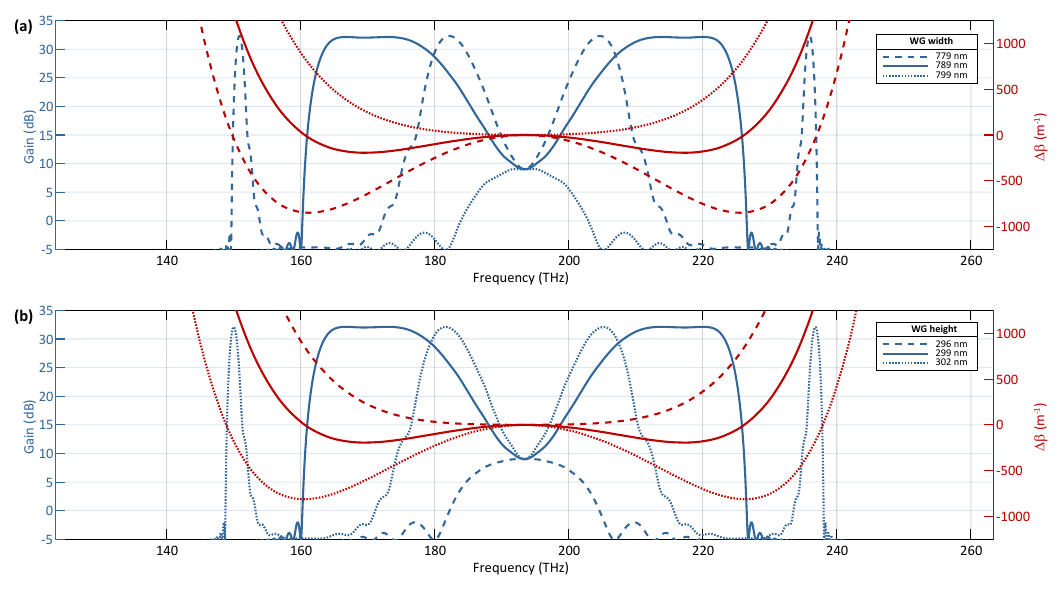}
	\caption{\textbf{Variations in gain spectra and propagation mismatch due to fabrication tolerances.}
		\textbf{(a)} Waveguide height is 299~nm, with waveguide width varied by $\pm$10~nm.
		Wider waveguide has normal dispersion and therefore significantly reduced gain.
		\textbf{(b)} Waveguide width is 789~nm, with waveguide height varied by $\pm$3~nm.
		A waveguide with a smaller height exhibits normal dispersion.
		The sensitivity of the dispersion profile is more pronounced compared to variations in the waveguide width.
		Variations in the cross-section that result in more anomalous dispersion do not notably affect the effective nonlinearity and the maximum gain but only lead to a reduction in bandwidth.}
	\label{fig:tolerance}
\end{figure*}

Numerical calculations of the signal gain $G_\textrm{S}$ and idler conversion efficiency $G_\textrm{I}$ are performed using frequency-domain nonlinear coupled-mode equations of the complex normalized amplitudes $A_\textrm{P,S,I} = \sqrt{P_\textrm{P,S,I}}$ of pump, signal, and idler waves\cite{Hansryd:02}
\begin{align}
	\frac{dA_\textrm{P}}{dz} &= \left( i\gamma \left(\vert A_\textrm{P} \vert^2 + 2\vert A_\textrm{S} \vert^2 + 2 \vert A_\textrm{I} \vert^2 \right) - \alpha/2 \right) \cdot A_\textrm{P} + i\gamma A_\textrm{P}^{\star} A_\textrm{S} A_\textrm{I} \cdot e^{i\Delta\beta z}, \notag \\
	\frac{dA_\textrm{S}}{dz} &= \left( i\gamma \left(2 \vert A_\textrm{P} \vert^2 + \vert A_\textrm{S} \vert^2 + 2 \vert A_\textrm{I} \vert^2 \right) - \alpha/2 \right) \cdot A_\textrm{S} + i\gamma A_\textrm{P}^2 A_\textrm{I}^{\star} \cdot e^{-i\Delta\beta z}, \notag \\
	\frac{dA_\textrm{I}}{dz} &= \left( i\gamma \left(2 \vert A_\textrm{P} \vert^2 + 2\vert A_\textrm{S} \vert^2 + \vert A_\textrm{I} \vert^2 \right) - \alpha/2 \right) \cdot A_\textrm{I} + i\gamma A_\textrm{P}^2 A_\textrm{S}^{\star} \cdot e^{-i\Delta\beta z},
	\label{eq:CoupledModes}
\end{align}
where $\alpha$ denotes the linear propagation loss and $\gamma$ denotes the effective nonlinearity of the GaP strip waveguide.
Here, $n_2$ is the nonlinear refractive index of the waveguide core and $A_\textrm{eff}$ is the effective mode area:
\begin{equation}
	\gamma = \dfrac{\omega_\text{P} n_2}{c A_\textrm{eff}}, \,\,\,\,
	A_\mathrm{eff} = \dfrac{\left(\int \vert\mathbf{E}\vert^2 dA\right)^2}{\int \vert\mathbf{E}\vert^4 dA}.
\end{equation}
All simulations are implemented via MATLAB (version 9.13.0 (R2022b)), while dispersion, effective nonlinearity and the effective mode area are calculated using COMSOL Multiphysics\textsuperscript{\textregistered}. 
For more precise simulations of broadband gain, the variation of mode profile with frequency should be taken into account; the effective mode area should be replaced with signal-frequency-dependent overlap integrals of the nonlinear interaction~\cite{agrawal2000nonlinear}.
For waveguides with a width of 790~nm and height of 299~nm, we find an effective mode area of 0.26~$\mu$m$^2$ and an effective nonlinearity $\gamma$ of 165~W$^{-1}$m$^{-1}$ at a pump wavelength of 1550~nm, with the nonlinear refractive index $n_2 = 1.1 \times 10^{-17}$~m$^2$~W$^{-1}$.
This value of $ \gamma $ is more than 300 times larger than that of dispersion-engineered \SiN\ waveguides~\cite{riemensberger2022photonic} and more than 10$^4$ times larger than typical highly nonlinear fibers~\cite{torounidis2006fiber}.
The nonlinear coupled-mode equations are integrated using a forward Euler scheme along the waveguide spiral of length $L$, and signal gain and idler conversion efficiency are calculated relative to the input signal power $P_\textrm{S}(0)$:
\begin{equation}
	G_\textrm{S}(L) = \dfrac{P_\textrm{S}(L)}{P_\textrm{S}(0)}, \,\,\,\,
	G_\textrm{I}(L) = \dfrac{P_\textrm{I}(L)}{P_\textrm{S}(0)} 
\end{equation}
In the small signal regime, the maximum gain in a waveguide of length $L$ is given by $ G_{\textrm{S}} = 1 + \left[ \gamma P_{\textrm{P}} g^{-1}\sinh(gL) \right]^2 $ and achieved when the linear propagation mismatch $ 	\Delta \beta = \beta(\omega_\textrm{S}) + \beta(\omega_\textrm{I}) - 2\beta(\omega_\textrm{P}) \approx \beta_2\left(\omega_\textrm{S}-\omega_\textrm{P}\right)^2 + \beta_4/12\left(\omega_\textrm{S}-\omega_\textrm{P}\right)^4 + \dots $ is compensated by the nonlinear phase-mismatch $ 2 \gamma P_{\textrm{P}} $ and, therefore, is defined by $ \Delta \beta + 2 \gamma P_{\textrm{P}} = 0 $ and $ g = \gamma P_{\textrm{P}} $, yielding $ G_{\textrm{S}} = 1 + \left[ \sinh(- \Delta \beta L/2) \right]^2 $.
Fig.~\ref{fig:gain_simulations}(a) shows simulated amplification spectra for the optimized waveguide cross-section.
Our devices turned out to feature a more strongly anomalous dispersion than simulated, resulting in the gain profile depicted in Fig.~\ref{fig:gain_simulations}(b).
This discrepancy may originate from imperfections and variations in the waveguide geometry and material stack that are not considered in the simulations.
The waveguide cross-section for the simulations in Fig.~\ref{fig:gain_simulations}(b) has been adjusted to match experimental results, giving a $\beta_2$ parameter of $-120$~fs$^2$mm$^{-1}$, while $\beta_4$ has a negligible contribution.
For the nominal cross-section of 789$\times$299~$\text{nm}^2 $, the theoretical value would be $\beta_2 =-16$~fs$^2$mm$^{-1}$ (and $\beta_4 = 3547$~fs$^4$mm$^{-1}$).
The experimentally measured value is $-124$~fs$^2$mm$^{-1}$ (Fig.~\ref{fig:dispersion_meas}), which is in a good agreement with numerical simulations for the adjusted amplification spectra, given that the spiral waveguide is short and precise measurements are challenging.
The value of $\beta_4$ cannot be reliable determined experimentally. 
We are currently performing additional measurements to determine the reason for the discrepancy between the experiment and simulations and to improve the material model of GaP, which is pivotal to achieve the full potential amplification bandwidth of 500~nm.
However, we are ultimately limited by fabrication tolerances.
	Fig.~\ref{fig:tolerance} shows that the parametric gain spectra can vary significantly if the cross-section of the fabricated waveguide deviates only slightly from the design.
	While variations of a few nm in waveguide width may be acceptable, a deviation in height of only 3~nm changes the gain profile dramatically.
In the literature, there is large uncertainty in the value of the nonlinear refractive index of GaP, ranging from $n_2 = 2.5 \times 10^{-18}$~W$^{-1} $m$^2$~\cite{martin2018nonlinear} measured by FWM to $n_2 = 1.1 \times 10^{-17}$~W$^{-1} $m$^2$~\cite{wilson2020integrated} measured via the optical parametric oscillation threshold in GaP ring resonators as well as by modulation transfer experiments.
We find that by using the smaller value, we underestimate the observed gain substantially.
Using the larger value, we need to reduce the input power in the simulation by 3~dB.
A similarly reduced Kerr parametric gain was recently observed in \SiN-based waveguide spirals~\cite{ye2021overcoming,riemensberger2022photonic}. 
The first reason for this observation may be residual higher-order nonlinear absorption in the GaP waveguide, as we operate the amplifier at high input power and close to the three-photon absorption threshold.
Moreover, Ye et al. have proposed that the reduced effective pump power stems from parasitic power transfer between different waveguide modes~\cite{ye2021overcoming}.
This is corroborated by the observation of substantial modulation of the transmission spectrum due to chaotic interference of the fundamental and higher order modes (see Fig.~\ref{fig:transmission}(a) and the discussion below), proving that at least some of the optical power is propagating in the higher order mode. 
Overall, we find that our observations support a value for the nonlinear refractive index towards the upper limit of the literature range.

\section{Transmission characterization and loss measurements}

\begin{figure*}[h]
	\centering
	\includegraphics[width=1\textwidth]{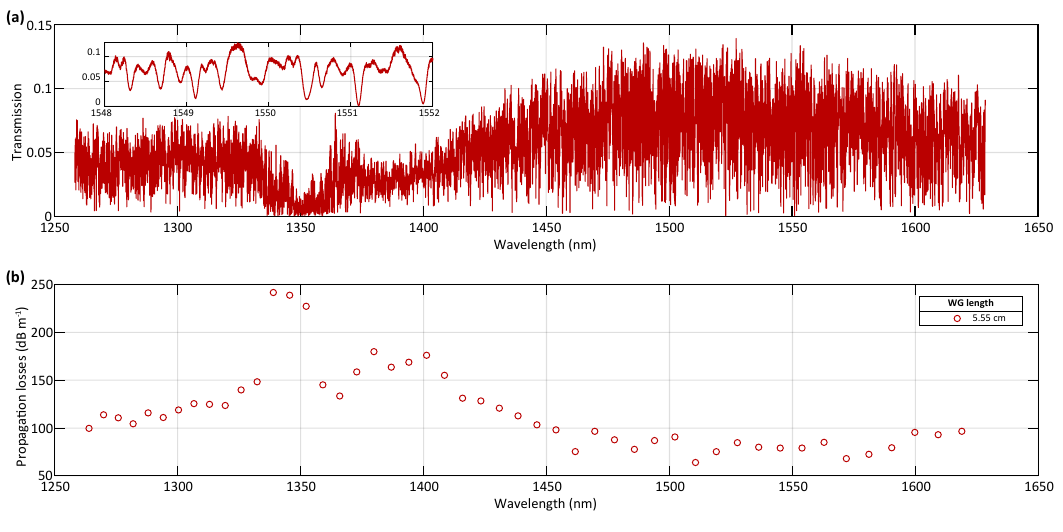}
	\caption{\textbf{Loss measurements of a GaP waveguide.}
		\textbf{(a)} Calibrated low-power transmission trace of the spiral waveguide with the waveguide width of 790~nm.
		The strong oscillations are attributed to higher-order mode interference, while the reduced transmission around 1350~nm is caused by hydrogen absorption.
		Inset: zoomed-in view of the region where the pump wavelength is typically set in our experiments.
		In certain regions, the total waveguide transmission can reach up to 12\%.
		\textbf{(b)} OFDR measurements of the same device, indicating frequency-dependent propagation losses.
		Increased losses at 1350~nm are consistent with previous observations of reduced transmission around this wavelength.}
	\label{fig:transmission}
\end{figure*}

The dispersion profile and propagation loss of the spiral are measured with a custom, all-band frequency comb-calibrated scanning diode laser spectrometer and frequency-domain reflectometer (OFDR)~\cite{soller2005high} over the wavelength range from 1260~nm to 1630~nm~\cite{DelHaye:09,riemensberger2022photonic}.
We characterize the transmission of our samples at low power, as described in~\cite{riemensberger2022photonic}.
A fully calibrated transmission trace of a 5.55-cm-long spiral waveguide is shown in Fig.~\ref{fig:transmission}(a).
The transmission exhibits a strongly oscillating behavior, that we attribute to the interference of the facet reflections and multimode interactions in the waveguide.
At wavelengths around 1550~nm, the overall transmission is 12\%.
However, during OPA gain measurements, the photonic chip is exposed to optical powers as high as 4.4~W, and the transmission trace is thermally red-shifted.
To maintain a good coupling of the pump laser when increasing the power, we adjust the pump wavelength, increasing it typically by not more than 1.6~nm at the highest available level of pump power.
In the S-, C- and L-bands the average loss rate is 0.8~dB~cm$^{-1}$ (Fig.~\ref{fig:transmission}(b)).
At shorter wavelengths, we observe increased losses that we attribute to absorption by the first overtone of the O-H stretch vibration in the low-temperature oxide cladding, which can be mitigated by using
different fabrication techniques~\cite{qiu2023hydrogen}.
From these measurements, we estimate an average coupling loss rate as high as 2.5~dB per facet, assuming that the input and output facets are the same.

\section{Optical gain measurements}

We use two widely tunable external-cavity diode lasers (ECDLs, TOPTICA CTL) as pump and signal sources.
The pump laser is amplified using an EDFA (Keopsys CEFA-C) up to 4~W.
The amplified spontaneous emission (ASE) from the EDFA is filtered out with a tunable bandpass filter (Agiltron FOTF), and the input power is controlled with a variable optical attenuator (Sch\"{a}fter+Kirchhoff 48AT-0).
One percent of each of the input waves is guided to power meters (Thorlabs S144C), and the rest is combined on a 10/90 fiber splitter, with the pump being injected in the 90~\% input.
We utilize lensed fibers to couple light into and out of the waveguide.
Of the collected light, 90~\% is guided to the power meter and 10~\% is analyzed using the optical spectrum analyzer (OSA, Yokogawa AQ6370D).
All input (output) powers quoted in the figures and the text are calibrated and indicate the values at the input (output) lensed fiber tips, unless specified separately.
The pump wavelength is set to 1550~nm.
For each pump power level, we continuously scan the signal laser wavelength from 1550~nm to 1630~nm (which is the maximum available wavelength for the laser that we use) while simultaneously scanning the OSA using the "Max Hold" function; at every new scan, the OSA records and updates only the highest values across the measurement span, while the signal laser wavelength is slowly swept to cover the entire amplification bandwidth.
To ensure that measurements at different wavelengths are performed under the same conditions without significant coupling degradation during the experiment, we slow down the laser scan speed and set the OSA resolution bandwidth to 2~nm. 
We find that in the optical amplification experiments, it is paramount to ensure good thermal coupling between the photonic chip and the chip holder to avoid excessive heating of the waveguide that leads to device failure and burning of the waveguide in the inverse taper section even at input powers as low as 1~W. 
We attribute this behavior to the fact that the pump wavelength is very close to the three-photon absorption threshold ($\approx$ 1660~nm for $E_\mathrm{g} = 2.24$~eV~\cite{lide2004crc}) and that therefore the nonlinear absorption of GaP increases strongly with temperature in the telecom wavelength region. 
Similar behaviour was observed in silicon~\cite{sinclair2019temperature} at telecom wavelengths and for GaP close the two-photon absorption threshold~\cite{ye2017optical}.
However, by ensuring good thermal contact between the chip and metallic chipholder, we can operate the OPA at room temperature and at 4.1~W input power for hours in air without active cooling.

\section{Parametric gain in waveguides with varying lengths}

\begin{figure*}[h]
	\centering
	\includegraphics[width=1\textwidth]{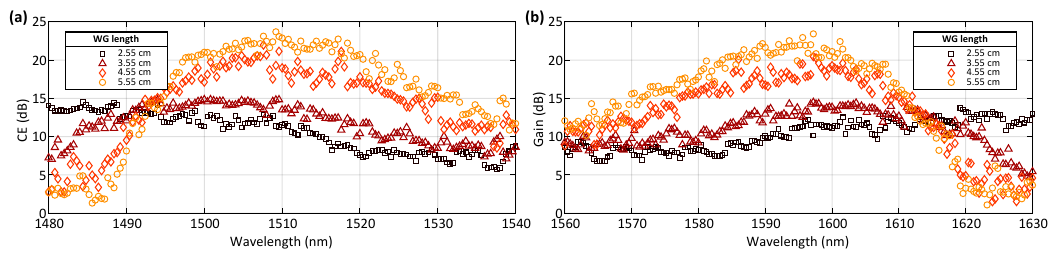}
	\caption{\textbf{Parametric amplification and idler conversion efficiency as a function of waveguide length}.
		\textbf{(a)} Idler conversion efficiency in waveguides with lengths ranging from 2.55~cm (black squares) to 5.55~cm (orange circles), recorded with a pump power of 3~W.
		All waveguides have cross sections of 780~nm $\times$ 299~nm.
		\textbf{(b)} Parametric amplification spectra for the same waveguides.
		Both conversion efficiency and parametric gain increase with waveguide length, while the bandwidth decreases due to accumulated phase mismatch.}
	\label{fig:length_sweep}
\end{figure*}

In addition to the continuous-wave amplification experiments presented in the main manuscript, we explored the behavior of the parametric gain and the conversion efficiency in spiral waveguides with the same design cross-section (waveguide width 780~nm), but different waveguide lengths, varying the length in steps of 1~cm from 2.55~cm to 5.55~cm.
The conversion efficiency and parametric gain data are depicted in Extended~Data~Figs.~\ref{fig:length_sweep}(A,~B), respectively.
Here, the pump power is set to 3~W.
As expected, the amplification bandwidth decreases with increasing waveguide length due to the accumulated linear phase mismatch, in good agreement with established theory~\cite{Hansryd:02} and previous observations~\cite{ayan2023towards}.
For the shortest length, 2.55~cm, we find that the net amplification bandwidth extends substantially beyond the tuning range of our laser.

\section{Spontaneous parametric comb formation}

\begin{figure*}[h]
	\centering
	\includegraphics[width=1\textwidth]{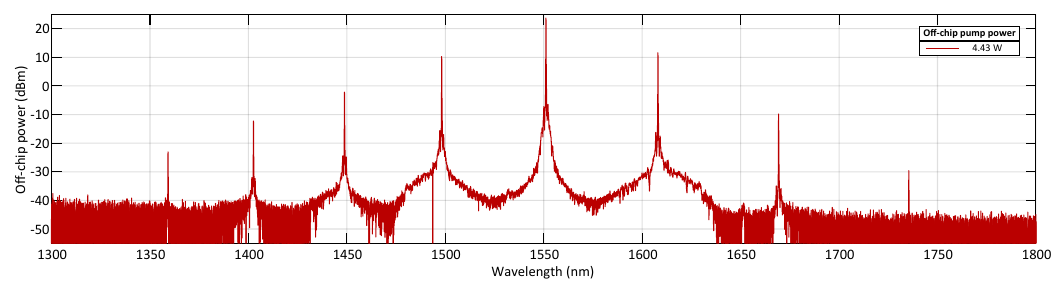}
	\caption{\textbf{Parametric comb formation in the waveguide with only the pump wave injected.}
		The pump power is set to 4.43 W; the positions of the first sidebands correspond to the maxima of the gain lobes.
		The spectrum is stable as long as the pump power is maintained high enough and the fiber-to-chip coupling is optimized.
		A small feature at 1650~nm indicates the Raman effect.
		Since the data was captured using a different OSA (Yokogawa AQ6375), the noise floor appears different compared to the spectra presented elsewhere in this work.}
	\label{fig:comb_formation}
\end{figure*}

In doped waveguide and fiber lasers, the achievable amplification gain is limited by spontaneous lasing through parasitic reflections from chip facets and other components. 
For example in recent work on erbium-doped waveguide amplifiers~\cite{liu2022photonic}, an off-chip net gain of 26~dB was achieved but required the use of index matching gel at the fiber facet to suppress parasitic lasing. 
In contrast, the strong yet unidirectional gain of the GaP TWPA allows us to achieve the same net-gain with simple cleaved chip facets and lensed fibers without parametric lasing because the threshold for the spontaneous sideband formation is intrinsically large compared to doped amplifiers with bidirectional gain.
However, operating the amplifier at 4.43~W, we occasionally observe spontaneous parametric oscillations in the waveguide spiral (Fig.~\ref{fig:comb_formation}), even in the absence of any input signal.
Due to the finite reflections from chip facets, amplified noise photons induce a coherent build-up, forming strong waves located at the maxima of the parametric gain lobes.
A similar behavior was recently observed in an optical fiber with single-sided Bragg reflectors~\cite{tovar2023random} due to Rayleigh scattering in the fiber.
Once formed, the comb can be present for a few minutes and disappears only when the coupling degrades too much.
A more rigorous investigation of the observed comb formation is out of the scope of this work and will be reported elsewhere.
This measurement demonstrates that our amplifier operates at or close to its internal gain limit, yet the achievable gain may be substantially improved if facet reflection is suppressed with the same methods as utilized in Ref.~\cite{liu2022photonic}.

\section{Signal power sweep and noise figure measurements}

\begin{figure*}[h]
	\centering
	\includegraphics[width=1\textwidth]{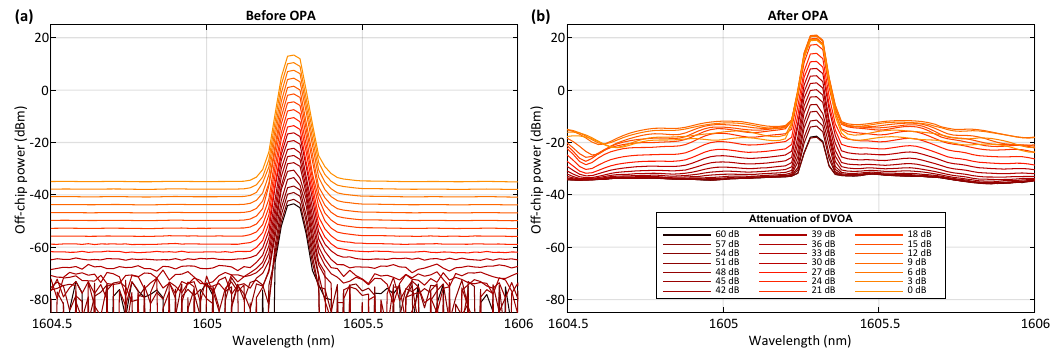}
	\caption{\textbf{Amplification measurements with varying input signal power.}
		\textbf{(a)} Optical spectra before amplification.
		The signal wavelength is chosen within the region of good transmission at the maximum of the parametric gain lobe to achieve the highest gain.
		The input power is swept from 46~nW (-43.4~dBm) to 21.9~mW (13.4~dBm) in steps of 3~dB.
		The optical SNR of the input signal exceeds 48~dB.
		\textbf{(b)} Optical spectra after amplification.
		An increased noise floor due to the parametric fluorescence is observed.
		As attenuation approaches 0~dB, the peak power exhibits only small changes, indicating amplifier saturation.}
	\label{fig:signal_power_sweep}
\end{figure*}

To vary input signal power, we use an L-band EDFA (Keopsys CEFA-L) and a digital variable optical attenuator (DVOA, OZ OPTICS DA-100).
These two components are installed in the signal optical path and used only for this experiment.
We set the signal EDFA to the maximum achievable output power, and we change the attenuation of the DVOA over the entire accessible range, from 60~dB to 0~dB.
For reference, we measure input signal power in the optical fiber before the OPA directly using the OSA in the absence of the pump wave (Fig.~\ref{fig:signal_power_sweep}(a)).
The amplified signal at a pump power of 4.43~W is shown in Fig.~\ref{fig:signal_power_sweep}(b).
We estimate the noise figure using the following relation~\cite{baney2000theory, ye2021overcoming}:
\begin{equation}
	NF = \dfrac{1}{G} + 2 \dfrac{\rho_{\textrm{ASE}}}{G h \nu},
\end{equation}
where $ G $ is optical gain and $ \rho_{\textrm{ASE}} $ is the noise power density, assuming a bandwidth of 0.1~nm.
We note that we use a corrected bandwidth for the measurements of the noise power density, and we account for 2.5~dB of coupling losses at each chip facet.

The optical SNR of the input signal is larger than 48~dB. 
Hence the contribution of the input ASE noise of the signal is negligible for small input signal powers, which can be seen in Fig.~\ref{fig:signal_power_sweep}(b); the noise power levels for the first input signal powers are essentially the same, making it possible to directly use the corresponding value as a true value for the noise power and eliminating the need to manually correct for the input signal ASE.
The signal noise here originates mostly from the L-band EDFA, and noise correction at high input signal powers is tricky because of the coupling fluctuations, spectral features in the transmission function of the device, higher-order FWM processes and gain saturation leading to different gains experienced by the signal and its noise.

\section{Frequency comb amplification -- extended data}

\begin{figure*}[h]
	\centering
	\includegraphics[width=1\textwidth]{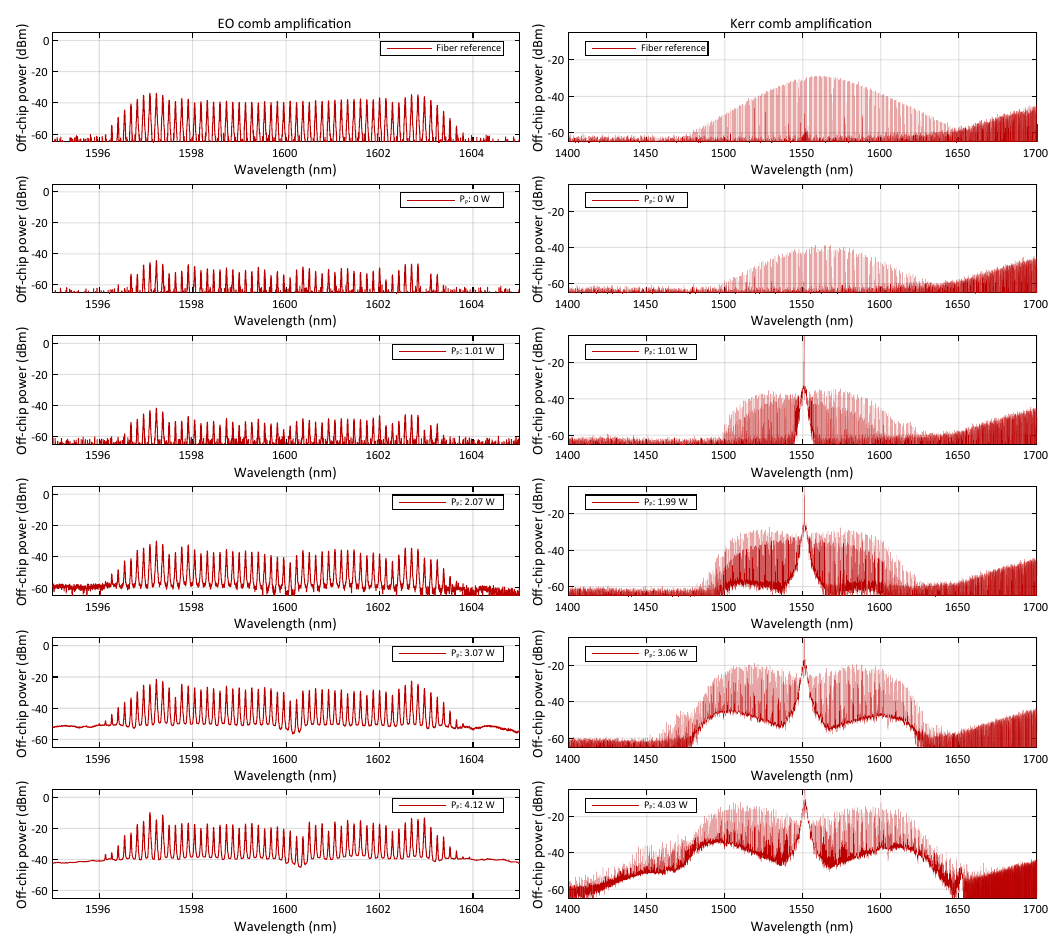}
	\caption{\textbf{Extended amplification measurement datasets for the EO comb and Kerr comb.}
		Here, P$_{\textrm{P}} $ is an off-chip pump power.
		Left column: EO comb amplification.
		The amplification is relatively uniform at the peak of the parametric gain lobe, with each line of the EO comb being amplified by approximately the same amount.
		The dip in the spectrum at wavelengths slightly above 1600~nm is caused by an increased loss in this spectral region, likely due to a defect in the waveguide or interference with higher-order modes.
		Right column: Kerr comb amplification.
		The enhanced signal comb overlaps with the generated idler comb.
		At the highest pump power, two distinct spectral features can be observed: additional lines from non-degenerate FWM, and a small peak around 1650~nm attributed to the Raman effect.
		The elevated noise floor at longer wavelengths is due to the roll-off of the OSA used to measure the spectra.}
	\label{fig:combs_extended}
\end{figure*}

In Fig.~\ref{fig:combs_extended}, we show the extended datasets for the EO comb and Kerr soliton comb amplification measurements.
Since we tune the pump wavelength at each pump power level to match the thermal drift of the spiral and to maintain good coupling, in the case of pumping with the Kerr soliton comb, the idler comb has a different frequency offset every time.
The idler comb can be seen in the data for the pump powers of 1.01~W and 1.99~W, while for higher powers, the offset between the OPA pump and soliton pump is almost a multiple of the comb FSR, and idler comb lines are located close to the signal comb lines.

\section{Coherent communication}

We use four different ECDLs (TOPTICA CTL) as sources of emission in this experiment -- pump, signal, and two local oscillators (LO) near the signal and idler wavelength -- to perform heterodyne measurements.
The QPSK modulation format at 10~GBaud symbol rate is used to encode a pseudo-random bit sequence in the signal using an IQ modulator (iXblue MXIQER-LN-30) driven by an AWG (Keysight M8195A) and a bias controller (iXblue MBC-IQ-LAB-A1).
Next, the modulated signal is transmitted to the OPA system where it is combined with a pump on a 10/90 fiber splitter, attenuating the signal by 10~dB.
Just before the OPA, the signal power in the fiber measures 0.5~$\mu$W. We direct 10~\% of the OPA output to the OSA, while the remainder proceeds along the primary measurement path.
We use a pair of filters, each containing a circulator and a fiber Bragg grating (FBG), to eliminate the residual pump in the fiber after the OPA section.
These filters allow only the signal and idler to advance through the communication line, attenuated by merely 4~dB as they pass through.
We use a C+L edge-band wavelength-division multiplexer (WDM) with a suppression ratio of more than 25~dB to split the signal and idler; we separately verified the normal operation of our WDM in the S-band.
To perform measurements using a coherent receiver (Finisar CPRV2222A-LP installed in Eva-Kit CPRV2xxx) we switch between signal and idler by simply changing the LO source and plugging the corresponding fiber output of the WDM into the input of the receiver.
We select the signal wavelength to be within the region of maximum gain and to be compatible with our equipment's capabilities.
The LO power levels are set to the maximum output of their respective lasers, after confirming that these values fall within the coherent receiver's specified input range.
The analog signal from the coherent receiver is collected via a fast oscilloscope (Teledyne LeCroy SDA8Zi-A) at a sampling rate of 40~Gs~s$^{-1}$ (i.e., 4 samples per symbol), and the obtained data is digitally processed using MATLAB Communications Toolbox (Version 7.8 (R2022b)) and custom functions.
The names of the functions are specified in parentheses below.
First, the imbalance in the collected IQ data is compensated using the Gram-Schmidt orthogonalization procedure~\cite{fatadin2008compensation} to ensure that the I and Q components of the signal are orthogonal, zero-mean, and normalized.
Next, we apply the coarse frequency offset compensation (\textit{comm.CoarseFrequencyCompensator}), and we use the automatic gain control (\textit{comm.AGC}) to remove amplitude variations.
The signal is decimated using a raised-cosine finite impulse response filter (\textit{comm.RaisedCosineReceiveFilter}) with the decimation factor of 2.
After this step, the timing synchronization (\textit{comm.SymbolSynchronizer}) using the Gardner timing error detector is added to the data processing algorithm to correct timing errors, and the signal is decimated again, reaching 1 sample per symbol.
We use the carrier synchronizer algorithm (\textit{comm.CarrierSynchronizer}) to correct carrier frequency and phase offsets for accurate demodulation of the received signal.
As a last step, we normalize the signal and equalize it using decision feedback filtering (\textit{comm.DecisionFeedbackEqualizer}) with the constant modulus algorithm.
To gain an understanding of the quality of the obtained constellations, we evaluate the modulation error ratio (MER)~\cite{ETR290} defined as
\begin{equation}
	M E R=10 \cdot \log _{10}\left(\frac{\sum_{k=1}^N\left(I_k^2+Q_k^2\right)}{\sum_{k=1}^N\left(e_k\right)}\right),
\end{equation}
where $e_k=\left(I_k-\tilde{I}_k\right)^2+\left(Q_k-\tilde{Q}_k\right)^2$ is the squared amplitude of the error vector, $\tilde{I}_k$ and $\tilde{Q}_k$ are measured in-phase and quadrature components of symbol vectors, and $I_k$ and $Q_k$ are ideal reference values (\textit{comm.MER}).
The MER essentially measures the spread of the symbol points in a constellation clusters.
A wider spread results in a lower MER and lower signal quality.
In the absence of significant signal degradation, the average measured symbol vector for each constellation point should coincide with the ideal symbol vector.
In this case, MER becomes equivalent to the constellation signal-to-noise ratio (or symbol SNR), which is typically calculated in a similar way but uses the average symbol vectors instead of the ideal reference points.

\newpage
\bibliography{bibliography}